*A Soliton Model of the Electron with an internal Nonlinearity cancelling the de Broglie-Bohm Quantum Potential*

ROALD EKHOLDT[§]

*Abstract*
*The paper proposes an envelope soliton model of the electron that propagates as a protuberance on a fictitious waveguide, which acts as trajectory. The model is based on de Broglie's original electron wave-particle relativistic theory, and on the observation that the Klein-Gordon equation governs the propagation mode of a common microwave waveguide—with the Schrödinger equation as a low group velocity approximation. This analogy opened for practical physical models, including solitons. In the linear case a conceived corpuscle zigzags within the fictitious waveguide—a zigzagging that resembles Penrose's picture of Dirac's electron theory. The soliton envelope is defined by a special nonlinear version of the Schrödinger equation. The nonlinearity cancels the de Broglie/Bohm Quantum Potential of the envelope. The wavefunction is confined by the envelope, and thus the concept of the collapse of the wavefunction is eliminated. Since the electron model is based directly on the Special theory of relativity it also illustrates that theory. The corpuscle, incorporating spin and charge, which moves at the speed of light, needs further study relating it to the photon. A nonlinear model of the photon, based on mutual coupling of two modes of an optical fiber at the Planck scale, is suggested. One of these modes is thought to carry two orthogonal polarized electromagnetic external fields, while the other, internal, mode carries a corpuscle that moves helically and thus represents the spin. A single photon is considered to have one linear polarization only, without any possibility for a circular polarization; hence, the concept of collapse in a polarization measurement is eliminated.*

*1. Introduction*

This paper is written in the spirit of Einstein's consoling words to de Broglie in Brussels in 1927 after the Solvay congress, where he was ridiculed by Wolfgang Pauli: *"Continue! It's you who are on the right course. […] a physical theory should, except for all calculations, be possible to illustrate by pictures so simple that a child should be able to understand them."* [2] But de Broglie gave up his project, until his Pilot wave idea was revived, independently, by David Bohm in 1952. [3] The following quotation from Bell predicts the soliton proposal remarkably well: *"The pragmatic attitude [ie following the Copenhagen interpretation], because of its great success and immense continuing fruitfulness, must be held in high respect. Moreover it seems to me that in the course of time one may find that because of technical pragmatic progress the 'Problem of Interpretation of Quantum Mechanics' has been encircled. And the solution, invisible from the front, may be seen from the back. For the present, the problem is there, and some of us will not be able to resist paying attention to it. The nonlinear Schrödinger equation seems to me to be the best hope for a precisely formulated theory which is very close to the pragmatic version."* [5] (He might have added the utility of Bohr's original hydrogen theory as well.) Moreover, Penrose writes: 'S*ome people object to this [introduction of nonlinearity], quite rightly pointing out that much of the profound mathematical elegance of standard quantum theory results from its linearity. However, I feel that it would be surprising if quantum theory were not to undergo some fundamental change in the future—to something for which this linearity would be only an approximation.*' [18]

The paper proposes an envelope soliton as a particle model of the electron. The soliton, which is of the spatial type, may be viewed as a protuberance on an otherwise fictitious waveguide, which acts as a trajectory. In the nonlinear case the wave and the corpuscle are confined to the soliton, while in the linear case both may be considered as propagating within the waveguide. These ideas are based on the observation that the Klein-Gordon equation governs the propagation mode of a common microwave waveguide—with the Schrödinger equation as a low group velocity approximation. This analogy opens for practical physical models—and by including non-linearity, solitons. The waveguide concept thus follows directly from de Broglie's original proposal, in his doctoral thesis, of the wave attributes of quantum mechanics, while the soliton concept relates to his Pilot wave and Double solution ideas. [1] Unfortunately, these ideas probably did not come to fruition as a model because they arrived before the microwave waveguide was invented in the nineteen-thirties, and before the envelope soliton theory was developed in the seventies. While the width of the waveguide in free space is determined by the reciprocal of $mc^2$, the width along the waveguide varies with the potential. The envelope is governed by a special nonlinear Schrödinger equation, which contains potential terms. By equating the nonlinear term with the de Broglie/Bohm Quantum Potential, the classical mechanics equation remains; ie as long as there are no abrupt variations of the potential. While the wave at first is seen as confined by the waveguide, it is finally seen as confined by the soliton envelope. Thus the concept of collapse of the wavefunction is eliminated.

[§] Retired—previously with The Norwegian Defence Research Institute
and the Norwegian Telecom Research Institute.
E-mail: rekholdt@online.no

The corpuscle is conceived as a photon-like entity, incorporating charge and spin attributes, which zigzags at the speed of light within the envelope—see figure 1.1. (Note that particle is used for the envelope soliton while corpuscle denotes the inner entity.) The charge and the magnetic dipole generate electromagnetic fields far outside the waveguide as well as coupling to such fields due to other sources. In this picture the Pauli exclusion principle may be considered to be due to direct coupling between two corpuscles. The zigzagging corresponds to Schrödinger's 'zitterbewegung' phenomenon, as well as Penrose's zigzag picture of the Dirac electron. [9] In the case of a potential barrier, the phase of the zigzag motion of the corpuscle will determine whether it, and thus the soliton, is reflected or not. As this phase is not amenable to measurement, at least with existing technology, it qualifies for the term hidden. This variable causes the fundamental probabilistic character of quantum mechanics, albeit not in Born's sense.

Although this model is local, the problems regarding spin and polarization of photons remain. A resolution of these problems requires a model of the photon itself. Hoping that the waveguide concept may apply to photons as well, this possibility is touched upon, very speculatively, by introducing dielectric rod, or fiber, waveguides, possibly at the Planck scale—and their two lowest modes. While the electromagnetic fields and the polarization are thought to belong to the lowest mode, the spin is imagined to belong to the second mode. One of these modes is thought to carry two orthogonal polarized electromagnetic external fields, while the other, internal, mode carries a corpuscle that moves helically and thus represents the spin. A single photon is considered to have one linear polarization only—due to the assumed nonlinearity the concept of linear superposition, and thus circular polarization is not valid; hence, the concept of collapse in a polarization measurement is eliminated.

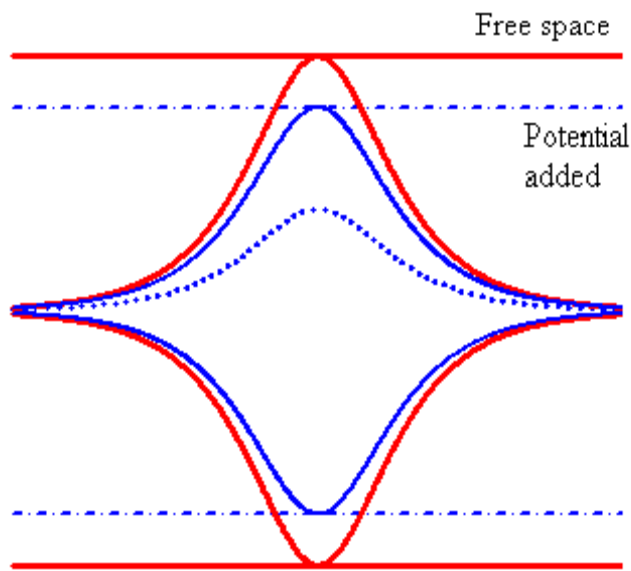

Figure 1.1. The electron as a spatial envelope soliton. The corpuscle and a standing wave oscillate within the envelope, which has a width determined by the rest energy and the potential.

*2. De Broglie's basic theory*
De Broglie took as his point of departure that quantum mechanics was based on the Planck-Einstein radiation quant, later called the photon, and that the Special theory of relativity played a decisive role. [7] Further, he believed that the electron and the photon were closely related, and that both had a wave-particle nature. He thus assumed that the electron's rest energy could be represented by $E_o = hf_o$ and momentum as $P = h/\lambda_o$ — where $f_o$ and $\lambda_o$ represent frequency and wavelength. But while the photon propagates at the velocity of light, the electron may even be at rest. To solve this problem de Broglie introduced the hypothesis that the electron, considered as a particle, had an inner vibration at a frequency that varied at the relativistic clock frequency:

$$f_{clock} = f_o \sqrt{1-(v/c)^2} \qquad (2.1)$$

where $v$ is the electron's velocity. Moreover, he assumed that a moving electron's energy varied relativistically as $E(v)= hf$, and thus that the frequency of the wave varied as

$$f_{wave} = f_o / \sqrt{1-(v/c)^2} \;. \qquad (2.2)$$

He thought that this wave could spread all over the physical, 3-dimensional, space, and even at higher velocities than the speed of light. But in his opinion this was not in conflict with the theory of relativity since the wave did not carry energy. This strange type of waves have quantum theory struggled with all the time afterwards. Schrödinger removed the particle all together, and only these waves remained.

Unfortunately, in view of his later insight into microwave waveguides, de Broglie could not at the time realize the following waveguide analogy; during the First World War he served at a military radio telegraph station at the Eiffel Tower. His co-worker and biographer Georges Lochak writes: *"From the war years he had a constant interest in the applications of physics, in particular the developments within radio technology, which he followed closely, to the latest progress in waveguides…".* [2] Actually, Feynman mentions this analogy briefly in the chapter on waveguides in his famous lecture series—but he does not seem to have elaborated on it. [8]

*3. The waveguide analogy*
Figure 3.1 depicts a topside view of a waveguide, similar to the common microwave rectangular metallic type used for the TE$_{10}$-mode, where we hypothetically assume a corpuscle and a piece of a wave-front zigzagging between the side-walls at the speed of light $c$. Interestingly, this figure resembles Penrose's zigzag picture of Dirac's electron theory. [9] *From his picture of the Higgs particle being related to the mass of every particle, we may picture the corresponding Higgs field as a pressure that acts on the waveguide walls.*

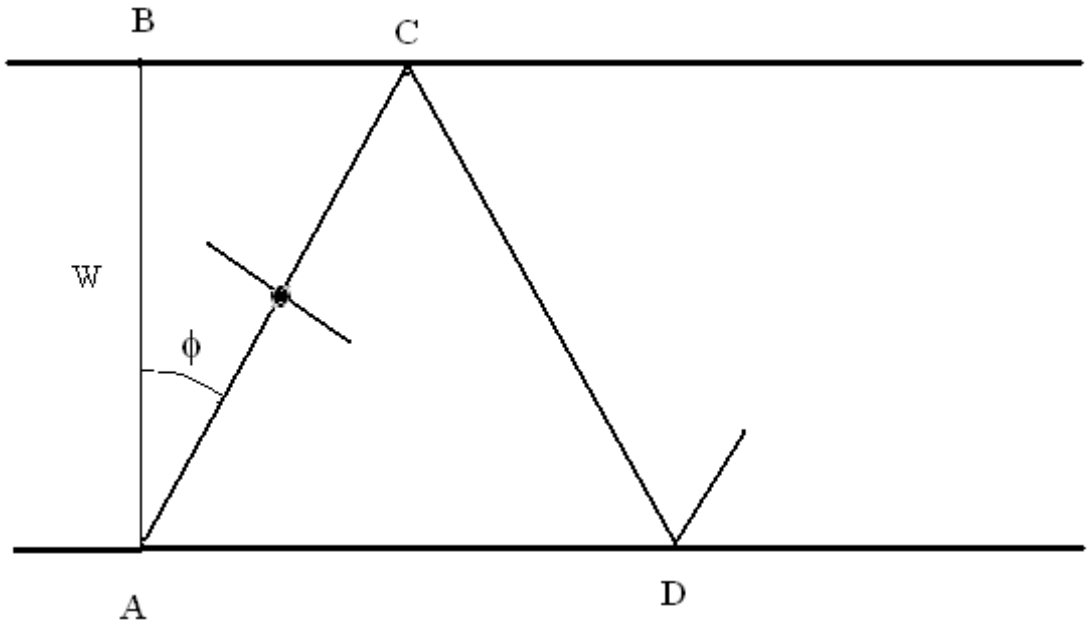

Figure 3.1. The corpuscle with a wavefront zigzagging between the waveguide's sidewalls.

The relation between the corpuscle's velocity along the waveguide and the angle $\phi$ is

$$v = c \sin\phi. \qquad (3.1)$$

Hence, the zigzag frequency is

$$f_{zigzag} = f_{clock} = f_o \sqrt{1-(v/c)^2} \qquad (3.2)$$

Here $v$ represents the effective propagation velocity of the signal's energy. From the triangle ABC in figure 3.1 we see that the phase velocity of the wave along the waveguide is

$$V_{phase} = c / \sin\phi. \qquad (3.3)$$

and thus that
$$vV_{phase} = c^2. \tag{3.4}$$

The cut-off frequency $f_o = m_o c^2 / h$ represents the lowest possible frequency for wave propagation—except for evanescent waves over short distances, equivalent to tunnelling in quantum mechanics. The corresponding waveguide dimension is thus $w = \dfrac{c}{2 f_o} = \dfrac{h}{2 m_o c} = 1.21 \times 10^{-11}$ m, which is one half the Compton wavelength, which is also approximately a quarter of the Bohr's radius for the lowest orbit of the hydrogen atom.

### *4. Similarity with Bohr's orbits in the hydrogen atom*

Since the details of de Broglie's development of Bohr's empirical quantization rule for the angular momentum in the hydrogen case largely appears to have been forgotten, it seems valuable to illustrate the waveguide analogy on that simple case—following Lochak's version, see appendix A. There we replace Bohr's ideal paths by waveguides. Here the width of the waveguide enters as an additional condition; hence, without making extra obscure assumptions, as Bohr had to do, we arrive at Bohr's non-relativistic, empirically based angular momentum relation. In conjunction with Bohr's model of the hydrogen atom, we note the analogy regarding the energy relation between the energy levels and the energy of corresponding photons and the frequency relations of frequency converters. This indicates the likelihood of a nonlinearity effect in the atom.

### *5. Wave equations*

De Broglie might have established a wave equation directly by using the relativistic energy relation
$E^2 = (m_o c^2)^2 + (cP)^2 = (hf_o)^2 + (ch/\lambda)^2$, where $P$ represents the linear momentum, and the corresponding dispersion relation

$$f^2 = f_o^2 + (ck)^2 ; \tag{5.1}$$

see figure 5.1. The corresponding wave equation is

$$\frac{\partial^2 \psi(z,t)}{\partial t^2} + \frac{\partial^2 \psi(z,t)}{\partial z^2} + f_o^2 \psi(z,t) = 0. \tag{5.2}$$

But before de Broglie, in parallel with others, in 1926 developed this so-called Klein-Gordon equation, Schrödinger in 1925 had successfully presented his non-relativistic equations. [10] He first tried the relativistic equation, but could not find agreement with empirical data, and did not publish his result. (His notes on this work seem to have disappeared when he escaped from Austria to Dublin in 1939.) By a rather *ad hoc* process he then developed his non-relativistic equations. The first was a time-independent equation for the hydrogen atom, where he combined standing waves, or vibrations, with classical mechanics as represented by the Hamilton-Jacobi equation. Schrödinger maintained the idea of standing waves even in his time-dependent equation, *which he introduced to account for time-varying potential energy, not as an equation of motion:*

$$i\hbar \frac{\partial \psi}{\partial t} + \frac{\hbar^2}{2m} \frac{\partial^2 \psi}{\partial z^2} - V(z)\psi = 0 \tag{5.3}$$

where $\hbar = h/2\pi$ and $V(z)$ represents the potential. The imaginary parameter $i$ in this equation is very strange.

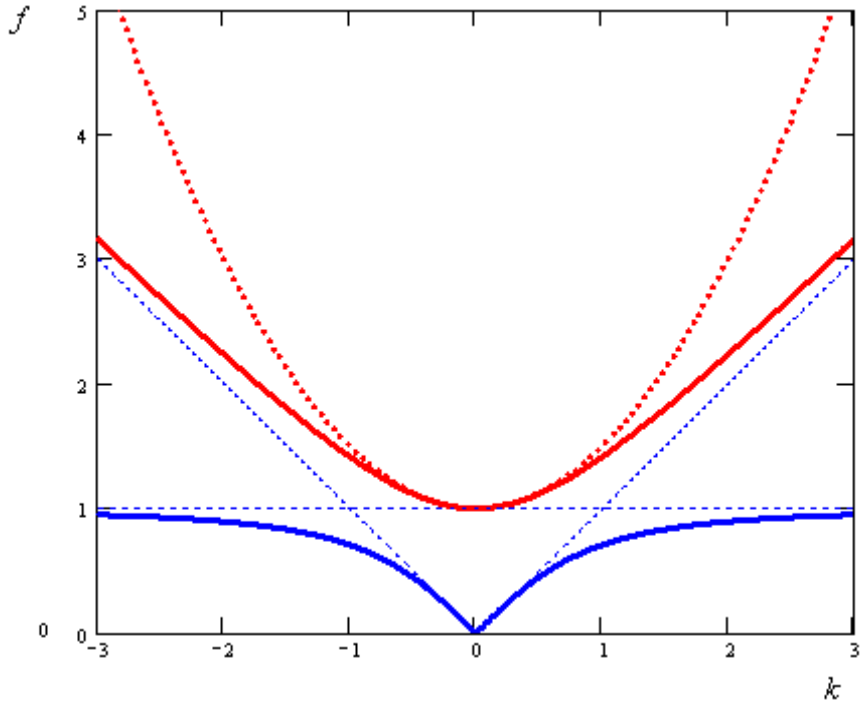

Figure 5.1. Dispersion relations, normalized to $f_o=1$ and $c=1$, corresponding to the Klein-Gordon, and its parabolic Schrödinger's approximation at low velocities—dotted. Note that the Schrödinger case here has the rest energy included—represented by $f_o=1$, but no frequency shift due to any potential energy. Note also that the lower clock frequency curve refers to $k_{clock}=1/k$, and that the normalized Compton wave number is one.

It is now usually overlooked that his intention was not that this equation should be an equation of motion for the electron as a particle. Strangely, de Broglie apparently did not pay attention to this, in our perspective, crucial difference. Although Schrödinger calls it the main difference, de Broglie does not mention it in the discussion of his reading of Schrödinger's papers, even if he is critical of three other aspects: *'J'ai lu à cette époque les mémoires de Schrödinger avec la plus vive admiration en réfléchissant beaucoup sur leur contenu. Sur trois points cependant, je ne me sentais pas d'accord avec l'éminent physicien autrichien. D'abord l'équation d'ondes qu'il attribuait à l'onde ψ n'était pas relativiste et j'étais trop convaincu de la liaison étroite existant entre la théorie de la Relativité et la Méchanique ondulatoire pour pouvoir me contenter d'une équation d'ondes non relativiste ; mais cette difficulté fut vite levée car, dès juilliet 1926, plusieurs auteurs, dont moi-même, ont trouvé une forme de l'équation d'ondes, connue aujourd'hui sous le nom d'équation de Klein-Gordon, dont l'équation de Schrödinger est la form dégénérée à l'approximation newtonienne. Un autre point où mes vues ne s'accordaient pas avec celles de Schrödinger était que celui-ci, tout en conservant l'idée que l'onde ψ dans l'espace physique est une onde réelle, semblait abandonner complètement l'idée de la localisation de la particule dans l'onde, ce qui ne concordait pas avec mes conceptions primitives. Enfin, tout en reconnaissant que la considération d'une onde ψ dans l'espace de configuration constituait un formalisme très utile pour la prévision des propriétés d'un ensemble de particules en interaction, je considérais comme certain que le mouvement des diverses particules et la propagation de leurs ondes s'opéraient dans l'espaces physique au cour du temps'.* [4] See English translation by the author in the footnote [**].

---

[**] 'At that epoch I read Schrödinger's papers with great admiration, while reflecting much on their contents. On three points, however, I did not feel to be in agreement with the eminent Austrian physicist. Primarily, the wave equation that he attributed to the wave ψ was not relativistic and I was too strongly convinced of the existence of a close connection between the Theory of Relativity and Wave Mechanics to be satisfied by a non-relativistic wave equation; but that difficulty was swiftly removed since, from July 1926, several authors, including myself, found a form of wave equation, which now is known as the Klein-Gordon Equation, of which Schrödinger's is a degenerate form that corresponds to the Newtonian approximation. Another point where my views were not in accord with Schrödinger's was that he, while retaining the idea that the wave ψ in the physical space is a real wave, seems to abandon completely the idea of the localisation of the particle in the wave, which was not in accordance with my initial conceptions. Finally, while recognising that considering a wave in the configuration space

Clearly, our model may be used to determine the bending of the waveguide in the case of transversal variation of the potential. As seen in equation (5.3), in addition to considering non-relativistic velocities only, the rest energy is omitted—which certainly was natural to do from a classical point of view; but thus the sidewalls of the waveguide were "blown" away. *Consequently, the whole universe was opened for the wave function, which in the waveguide model is confined to the inside of the guide. Thus the analogy with Bohr's orbital model of the hydrogen atom disappeared, which from this point of view created great problems with regard to the understanding of quantum mechanics. The most serious consequence was, however, that it required the problematic artificial concept of the collapse of the wavefunction to relate the wavefunction results to those of measurements.*

If we as an analogy consider $\omega_o$ as a carrier in a modulated transmission system, Schrödinger, by removing the carrier, shifted the signal down to baseband, but without demodulating to recover the modulating signal. In a transmission system one would mix the modulated carrier with the carrier in a nonlinearity and than filter away the remains of the carrier. By removing the carrier first, Schrödinger implicitly barred the interpretation of the wavefunction as a modulated carrier.

*6. The Born rule.*

We shall now take a look at the generally accepted statistical nature of quantum mechanics according to Max Born's rule. We approach this interpretation by revisiting the de Broglie's original analogy between the photon and the electron, as well as our corresponding waveguide analogy. Initially, we shall thus consider the density of the power of a continuous, monochromatic wave propagating in a microwave waveguide. This density varies proportionally with the inverse of the group velocity, which again varies with the width of the waveguide. Moreover, the power density is proportional to the density of photons that all correspond to a definite energy—ie frequency. Also, the power is proportional to the quadratic value of the field, which propagates as a wave.

Shifting to the electron, we consider an ensemble of equal, non-interacting fictitious electrons without spin. We anticipate a round-shaped pulse—which we shall return to in chapter 7. While in the photon case the group as well as the phase velocities are close to *c*, we consider here electrons in the non-relativistic case. Moreover, in the following we subtract the rest energy from the energies *E* and *V* to be in line with the normal terminology in quantum mechanics. From equation (5.5) we see that there is a frequency residue, but in the non-relativistic case this frequency is negligible. The remaining equation is thus

$$E = V(z) + \frac{(\hbar k)^2}{2m_o}, \quad (6.1)$$

and hence

$$k^2 = \frac{2m_o}{\hbar^2}[E - V(z)] \quad (6.2)$$

and consequently

$$v_{group} = \sqrt{2[E - V(z)]/m_o}. \quad (6.3)$$

The corresponding wave equation is

$$\frac{\partial^2}{\partial z^2}\psi + \frac{2m_o}{\hbar^2}[E - V(z)]\psi = 0 \quad (6.4)$$

Thus, the field has the structure

$$\psi(z) = A(z)exp\{+/-i\sqrt{2m_o[E - V(z)]}\, z/\hbar^2\} \quad (6.5)$$

The equation (6.4) is of course Schrödinger's time-independent wave equation. (Actually, Schrödinger admitted that the term wave-equation was misleading; he would have preferred vibration- or amplitude-equation. In this context it should again be emphasized that his intention of the time-dependent equation was to cover the case of time-dependent potential, not to introduce travelling waves—see [10]. From our point of view it was unfortunate that others interpreted his time-dependent equation as a travelling wave equation—and even as an

---

constitutes a very useful formalism for the prediction of the properties of an ensemble of interacting particles, I considered as certain that the movement of the diverse particles and their propagation operated in the physical space and time.'

equation of motion for probability waves. Thus, we may also in this case of a definite energy *E* consider the equation

$$\frac{\partial^2}{\partial z^2}\psi - \frac{2m_o}{\hbar^2}V(z)\psi = E\psi \tag{6.6}$$

This equation may be interpreted as an eigenvalue equation, and hence the method of eigenfunction expansion may be applied. *The definite energy condition corresponds to the state concept.* Since we are dealing with a linear equation we may generate the sum of two waves that move in opposite directions

$$\psi(z) = A(z)exp\{-i\sqrt{2m_o[E-V(z)]}\,z/\hbar^2\} + B(z)A(z)exp\{i\sqrt{2m_o[E-V(z)]}\,z/\hbar^2\}$$
(6.7)

where *A(z)* and *B(z)* are to be determined by boundary conditions.

Consequently, by this procedure we obtain the probability density for electrons, and we have thus reproduced the *ad hoc* Born rule of quantum mechanics: $P_d(z) = |\psi(z)|^2 / \int |\psi(z)|^2\,dz$.

## 7. Wave packet as an envelope soliton—a heuristic approach

By considering the de Broglie's wave-particle model we have so far, more or less tacitly, assumed a particle and, in principle, an unlimited continuous wave. Schrödinger considered representing the particle by a pulse, but Lorentz dissuaded him from it by arguing that the pulse would undergo dispersion. [10] Of course, they had no knowledge of the soliton concept, where nonlinearity fully compensates for dispersion.[††] In the nineteen-fifties de Broglie in his revived work on the Double solution concept introduced the need for nonlinearity, but this was ten years before the great development of nonlinear theory took off—especially by the introduction of the soliton concept, which was developed early in the seventies. Since we are dealing with a dispersive waveguide, an envelope solition solution seems to be an interesting idea for the wave-particle concept. To approach the soliton model in our waveguide setting we introduce what is called a potential well in quantum theory; ie we insert a broader waveguide section. We tentatively choose the width of the section to be twice the width of the ordinary waveguide—see figure 7.1. This choice is based on the half value of the spin of the electron. Moreover, this section is assumed exited by a wave at a lower frequency than the cut-off frequency of the ordinary waveguide, which thus is only entered by an evanescent wave. Here we assume the frequency to be the sub-harmonic of order one half of the cut-off frequency of the ordinary waveguide; the latter frequency may from our point of view be interpreted as a representing a universal field—a Higgs type of field? The generation of sub-harmonic is a relatively rare nonlinear effect—see [19].

The standard nonlinear Schrödinger (NLS) is considered as a generic equation for unidirectional propagation of wave packets in a dispersive medium at the lowest order of nonlinearity. [11] It is usually presented in normalized form:

$$i\frac{\partial u}{\partial t} + \frac{\partial^2 u}{\partial z^2} + 2|u|^2 u = 0 \tag{7.1}$$

The exact solution to this nonlinear equation is

$$u(z,t) = a \cdot exp[\,ivz/2 + i(a^2 - v^2/4)t\,] \cdot sech[\,a(z - vt - z_o)\,],$$

which is a so-called breather moving at velocity *v*.

For *v = 0* the solition is reduced to $u(z,t) = a \cdot exp(ia^2 t) \cdot sech(az)$

That the potential is only represented by the nonlinear term in the NLS, clearly rules out its direct application in quantum mechanics. This problem may be overcome in the Gross-Pitaevskii (GP) equation [12]:

$$i\hbar\frac{\partial \varphi}{\partial t} + \frac{\hbar^2}{2m_o}\frac{\partial^2 \varphi}{\partial z^2} - mc^2 - V\varphi - g|\varphi|^2 \varphi = 0, \tag{7.2}$$

where *g* may be positive or negative. Unlike the NLS, however, this equation does not have an exact solution; consequently, approximate solutions are needed.

---

[††] It is somewhat ironic that for the so-called envelope solitons—which are applied in optical fiber communication—the amplitude is described by a nonlinear version of the Schrödinger equation.

However, having so far followed de Broglie's remarkable intuitive approach towards the reality underlying quantum mechanics, we shall instead try his Pilot Wave idea from 1927—which he saw as a step towards the Double Solution.

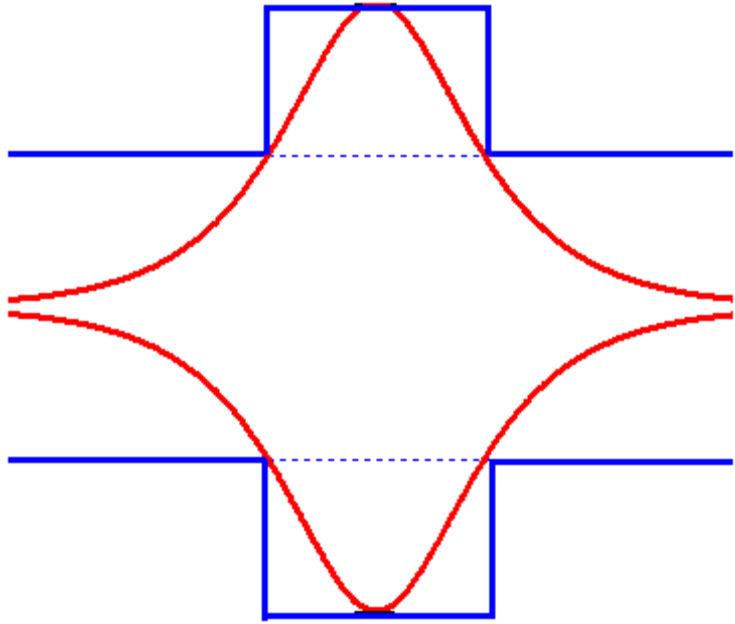

Figure 7.1. An assumed *sech* shaped soliton within an inserted, fictive waveguide section.

As this idea is better known as Bohm's Quantum Potential theory from 1952, we shall use his terminology. [3] Wishing to consider the classical limit in quantum theory, Bohm assumed the WBK approximation[‡‡], and introduced the wave function as

$$\psi = R \cdot exp(iS/\hbar),  \quad (7.3)$$

where $R$ and $S$ are real functions. By substitution into equation (5.3) and by separating the real and imaginary parts, we get the two following equations:

$$\frac{\partial S}{\partial t} + \frac{1}{2m_o}\left(\frac{\partial S}{\partial z}\right)^2 + V - \frac{\hbar^2}{2m_o}\left(\frac{\partial^2 R}{\partial z^2}\right)/R = 0 \quad (7.4)$$

$$\frac{\partial R}{\partial t} + \frac{1}{2m_o} \cdot \left[ R^2 \frac{\partial^2 S}{\partial z^2} - 2 \frac{\partial R}{\partial z}\frac{\partial S}{\partial z} \right] = 0 \quad (7.5)$$

Except for the last term, equation (7.4) is equal to the Hamilton-Jacobi equation in classical mechanics, where $S$ has the dimension of action,

$$\frac{\partial S}{\partial t} + \frac{1}{2m_o}\left(\frac{\partial S}{\partial z}\right)^2 + V = 0 \quad (7.6)$$

The equation (7.4) may alternatively be presented as

$$\frac{\partial S}{\partial t} + \frac{1}{2m}\left(\frac{\partial S}{\partial z}\right)^2 + mc^2 + V + Q = 0 \quad (7.7)$$

where

$$Q = -\frac{\hbar^2}{2m_o}\left(\frac{\partial^2 R}{\partial z^2}\right)/R \quad (7.8)$$

is called the Quantum Potential.

---

[‡‡] The Wentzel-Kramer-Brillouin approximation is based on the assumption of a slowly varying potential along the trajectory.

Tentatively we try the envelope to have the shape *sech( az )*—see figure 7.1. This shape seems, from our physical point of view, to match both a standing wave inside the inserted section as well as an evanescent wave entering the outside waveguide. From our waveguide perspective it seems interesting to picture an envelope solution as a further confinement of the wavefunction.

As a consequence of the choice

$$R = sech( az ) \qquad (7.9)$$

we may interpret the second derivative of $R$ as representing the frequency variation over the pulse; hence $Q$ represents the corresponding potential energy variation over the pulse. These variations thus appear to represent the dispersion. In this case

$$Q = -\frac{\hbar^2 a^2}{2m_o}\left[tanh^2( az ) - sech^2(az)\right] = -\frac{\hbar^2 a^2}{2m_o}\left[sinh^2( az ) - 1\right]sech^2(az) \qquad (7.10)$$

See figure 7.2. Thus, $-Q$ added on the left hand side in (7.7) cancels the dispersion, and this equation becomes equal to (7.6):

$$\frac{\partial S}{\partial t} + \frac{1}{2m_o}(\frac{\partial S}{\partial z})^2 + V = 0$$

Furthermore, the Schrödinger equation may be seen as getting a nonlinear term $-Q$:

$$i\hbar\frac{\partial \psi}{\partial t} + \frac{\hbar^2}{2m}\frac{\partial^2 \psi}{\partial z^2} - V( z )\psi + \frac{\hbar^2}{2m_o |\psi|}(\frac{\partial^2 |\psi|}{\partial z^2})\psi = 0 \qquad (7.11)$$

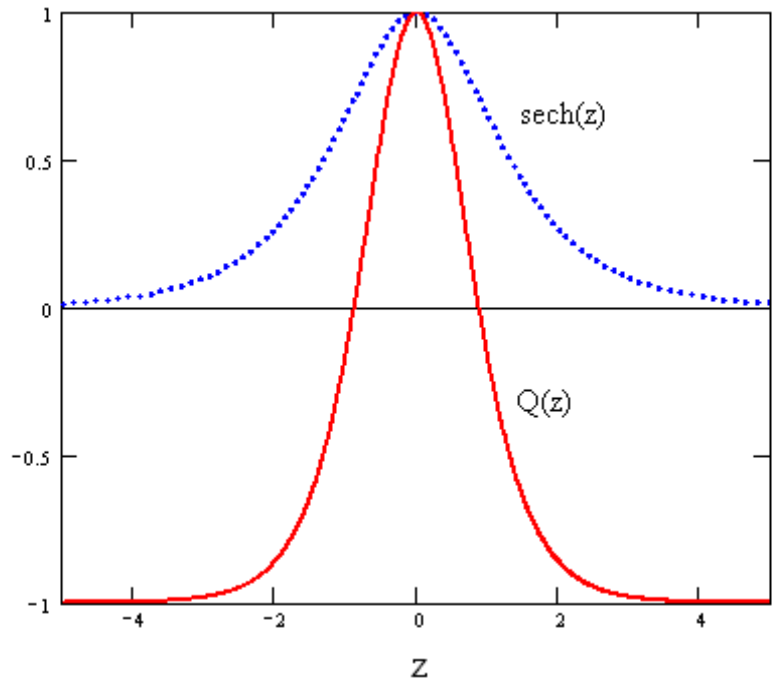

Figure 7.2. Quantum potential corresponding to *R=sech(az)*; (normalized).

The heads and tails of the soliton will only come into play at strong variations of the potential, especially square potential barriers—where these parts may contribute to non-classical tunnelling. Hence, when the soliton model impinges on such a barrier, the phase of the zigzag motion of the corpuscle may determine the outcome:
- If the corpuscle hits the barrier, it is reflected
- If it hits an opening in the barrier that continues into a narrower waveguide, it will adapt itself to this waveguide, and continue as in the WBK case, if that is valid.

While this outcome in principle is deterministic, to measure the zigzag phase is impossible and may thus be considered to be hidden. Furthermore, the relativistic case in free space may be obtained by using the classical relativistic transforms—since we have based our tentative theory on de Broglie's basic relativistic assumptions.

*8. Heisenberg's Uncertainty Relations*

Since the electron in our model is represented by an envelope in the shape of a pulse, one may tentatively explain the uncertainty relations by analogy with the Ambiguity function of simple pulse-Doppler radars—see the similarities in the following sequence:

$$\Delta E \Delta t > h/2$$
ie
$$\Delta f \Delta t > 1/2$$

Thus, high Doppler resolution requires long pulses, which impair distance resolution.

$$\Delta f \, v_g^{-1} \, v_g \Delta t > 1/2$$
ie
$$(\Delta \lambda)^{-1} \Delta x > 1/2$$
ie
$$\Delta k \Delta x > 1/2$$
ie
$$\Delta p \Delta x > h/2$$

This idea is not new, but has generally been rejected.

*9. The photon—a dual mode spatial envelope soliton?*

Our envelope soliton model of the electron, which is based on an analogy between its trajectory and a common microwave waveguide, makes it tempting to look for a similar wave-particle model of the photon based on a waveguide trajectory. Their differences, however, amount to serious obstacles, since the photon has

- No known mass.
- No charge.
- Fixed velocity *c* in free space.
- Frequency range down to zero
- Gravity influence

The following is a brief, speculative sketch of a waveguide based model of a localised photon.

The low frequency wave problem possibly indicates the lowest mode of a circular dielectric rod, or fiber, waveguide. This so-called N1 mode has zero as low frequency cut-off, and belongs to a family that has sinusoidal variations of fields about the circumference. Thus, the N1 mode has a single $2\pi$ sine variation. [17] At relatively low velocities this mode corresponds to electromagnetic vector fields, and waves. On the other hand, the corpuscle part of the photons may be seen as belonging to the N2 mode, which has a $4\pi$ sine variation. The index of refraction inside the rod is assumed to be only slightly larger than at the outside to open for total reflection.

Moreover, the influence by gravity on light, according to the General theory of relativity, may indicate that the Planck scale comes into play—for instance that the diameter of a straight rod be assumed to be of the order of the Planck length:

$$l_p = \sqrt{\hbar G / c^3}$$

Here *G* is the gravitational constant. This length is of the order of $10^{-35}$ meter, which would correspond to an extremely high cut-off frequency of the N2 mode. A gradient of the gravitational field across the rod would conceivably bend it.

At low frequencies the N1 mode would have an evanescent part extending far outside the rod. This feature may possibly explain the dual slit experiments: While the N1 mode extends over both slits, the N2 mode, being confined to the inside of the rod due to total reflections, passes only through one of the slits.

Figure 9.1 represents the dispersion curves of the two modes. The upper curve represents the N2 mode, with a lower curve representing the periodicity in time and length of a corpuscle travelling helically inside the rod. The N1 mode, which essentially follows an ideal non-dispersive velocity $c_o$, is assumed to match and to be mutually coupled to this helical "mode" at the low frequencies—which are the frequencies we attribute to the photons. In figure 9.2 these roundtrips are pictured as a ray in a quadratic shaped spiral inside the rod. Together the corpuscle and the pulse make up the photon. We may assume the spiralling corpuscle to travel at an effective velocity a little bit lower than $c_o$ inside the rod. The length of the photon, as a spatial envelope soliton, will thus be limited by the coherency conditions, ie on the relative phase of the corpuscle in the spiral and in the N1 pulse. With a sufficient number of periods with phase coherency of the modes the single photon may be amenable to the observed filtering that is equal to the filtering of the corresponding continuous wave.

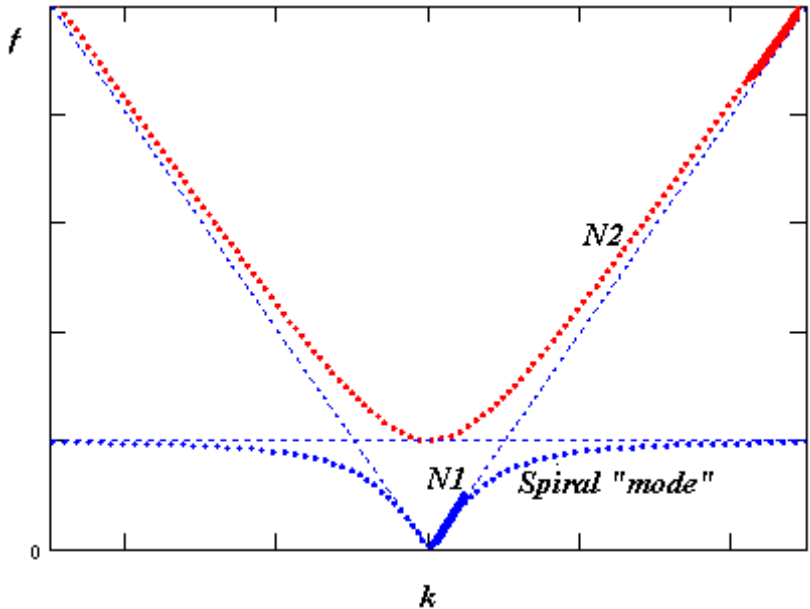

Figure 9.1. The upper curves represent the N2 mode with dispersion similar to the $TE_{10}$ - mode of the electron model. The lower curves may be seen as representing the N1 mode and the helix of the corpuscle in the N2 mode; the gradients of these lower curves correspond to the common group velocity of the coupled N1 and N2 modes. *The cut-off frequency point of the N2, which in analogy with the electron model should represent the rest energy, and hence the mass, is purely fictitious.* This is due to the total reflection assumption that restricts the validity of the dispersion curves to group velocities extremely close to the ideal velocity of light. The active parts of the modes are indicated by solid lines. <u>The figure is not to scale</u>.

*The cut-off frequency point of the N2, which in analogy with the electron model should represent the rest energy, and hence the mass, is purely fictitious—due to the total reflection assumption that restricts the validity of the dispersion curves to group velocities very close to the velocity of light. With the diameter of the rod of the order of the Planck length, the fictitious mass would be of the order of the Planck mass*

$$m_p = \sqrt{\hbar c / G}$$ *, which is of the order of 10$^{-8}$ kilo—which is just a little less then a flea!*

A small evanescent effect at the corners of the helical ray may act as a coupling to the N1 mode. As depicted in figure 8.2, the coupling between the N1 mode and the helical "mode" may thus be assumed to give rise to two orthogonal polarizations in the N1 mode. If that is the case, a photon may be thought always to belong to one of the polarizations. One of these modes is thought to carry two orthogonal polarized electromagnetic external fields, while the other, internal, mode carries a corpuscle that moves helically and thus represents the spin. A single electron is considered to have one linear polarization only—no circular polarization; hence, the concept of collapse in a polarization measurement is eliminated. This may be seen as a result of the assumed introduction of nonlinearity, which does not concur with the linear superposition concept. See [20]

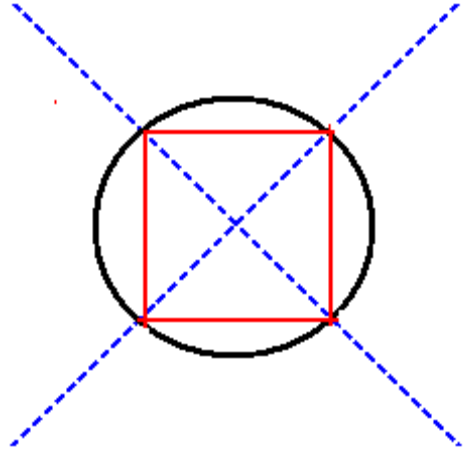

Figure 9.2. An optical ray approximation illustration of the N1 and N2 mutually coupled modes of a circular dielectric rod, where the inside quadratic helical trajectory of the corpuscle is projected onto a transversal fictional rest plane. While the N2 mode is confined to the rod, the evanescent part of the N1 mode may at low frequencies extend far outside the rod. The coupling between N1 and the "mode" at its corner may be assumed to create two orthogonally polarized solution of the N1 mode. The helicity +/- *h* of the photon thus only belongs to the N2 mode, not to the electromagnetic N1 mode.

It should be kept in mind that the mode coupling is mutual, which implies that external influences on the N1 mode may turn the N2 pattern and hence the polarization.

## *8. Concluding remarks*

Our hypothesis of a soliton wave-particle model of the electron following a waveguide trajectory has a reactionary character—as the situation before Schrödinger presented his equations is brought up. For instance, it revives the original Bohr-Sommerfeld classical atom model by explaining the roles of classical mechanics as well as of wave mechanic in the prescription of the discrete orbital structure. In general, the model reduces the width of the gap between classical and quantum mechanics. Compared to prevailing interpretations of quantum mechanics, the main differences are:

- The collapse of the wavefunction is eliminated, since the electron and its wavefunction are confined to the envelope soliton.
- The indeterminism, as a basic principle, is likewise eliminated—even if limited measurement possibilities prevent the resolution of the apparent statistical phenomena into elementary events. The hidden variable idea is thus revitalized (at least as far as the electron is concerned).
- The Born rule regarding a statistical interpretation is clarified.
- The Heisenberg uncertainty relations may be viewed as an analogue to the Ambiguity relation for pulse-Doppler radars. [15].
- Quantum Mechanics and the Special theory of relativity seem to be reconciled, since the waveguide model is directly based on the latter.

Although the electron model is local, the problems regarding spin and polarization of photons remain. A very speculative soliton model of the photon is suggested. This model is based on the two lowest modes of a dielectric rod—with the diameter at the Planck scale. The electromagnetic fields and their polarization are considered to belong to the lowest mode, while the spin belongs to the second mode. A single photon is considered to have one polarization only, without any possibility of circular polarization; hence, the concept of collapse in a polarization measurement is eliminated.

Coupling of two, or more, electrons has not been considered; those aspects clearly needs further studies—which would require a model of the photon-related corpuscle, where the spin and charge are assumed to be located.

*Appendix A*
*Similarity with Bohr's orbits in the hydrogen atom*
Since the details of de Broglie's development of Bohr's empiric quantization rule for the angular momentum in the hydrogen case appears to have been entirely forgotten, it seems valuable to illustrate our waveguide analogy on that simple case—following Lochak's rendition; see reference [7].

In his planetary model Bohr applied classical mechanics, where the centrifugal force balances the attractive force of the nucleus:

$$mv_e^2/r = e^2/r^2 \quad \text{i.e.} \quad mv_e^2 r = v_e M = e^2 \qquad (A.1)$$

where $M$ represents the angular momentum. The classical energy of the hydrogen atom is thus

$$E = \frac{mv_e^2}{2} - \frac{e^2}{r} = -\frac{mv_e^2}{2} \qquad (A.2)$$

From the empiric spectral lines Bohr established the relation $M = Nh/2\pi$, where $N$ is an integer quantum number.

*De Broglie based his treatment of the hydrogen atom on his Accordance of phases law, which he considered as his most important achievement in quantum theory*—reference 4, pages 3-4. He found that for a Galilean observer the phase of the internal vibration and the phase of the wave along the path were identical at any time:

$$\varphi_{clock}(z) = f_{clock} t = f_{clock} \frac{z}{v_e} \qquad (A.3)$$

and

$$\phi_{wave}(z) = f\gamma(t - V_{phase}\frac{z}{c^2}) = \frac{f_o}{\gamma}(\frac{z}{v_e} - \frac{v_e z}{c^2}) = f_o \gamma \frac{z}{v_e} = f_{clock} \frac{z}{v_e} = \phi_{clock}, \qquad (A.4)$$

where he applies the Lorentz transform for time.

*Again it seems unfortunate that he did not know the waveguide analogy, because the accordance of the two phases is evident from the dispersion relation, equation 5.6, and the triangle ACD in figure 3.1.*

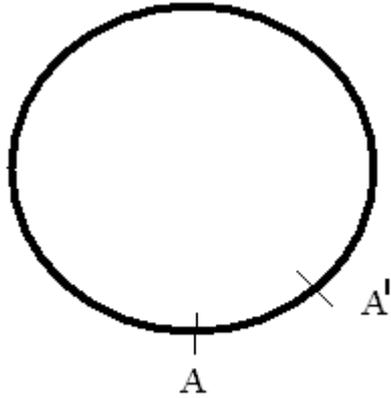

Figure A1. After the electron has made a full circle from the origin A, it has to move an extra arch A-A' for their joint equal phase at the origin of the wave and the zigzag movement to reappear.

De Broglie chose a point A, see figure A1, as a common origin for the electron and the wave to start at the same time and phase. When the particle has made a full circle at time $T$, it has to move an additional arch A - A' in an extra time $\tau$ due to the relativistic time dilatation. Thus, according to equation 3.4

$$V_{phase}\tau = c^2\tau/v_e = (\tau + T)v_e. \qquad (A.5)$$

Hence

$$\tau = \frac{v_e^2}{c^2 - v_e^2} T. \qquad (A.6)$$

De Broglie then supposed that stationarity required that the phase from A to A' had run through $2\pi N$ radians — where $N$ is an integer, ie

$$2\pi \cdot f_{clock} \tau = 2\pi \frac{m_o c^2}{h} \sqrt{1-(v_e/c)^2} \frac{v_e^2}{c^2 - v_e} = 2\pi N \tag{A.7}$$

This means that we have $m_o v_e^2 T / \sqrt{1-(v_e/c)^2} = Nh$ —or non-relativistically $m_o v_e^2 T = Nh$, which is Bohr's empirically based relation. That the wave pattern moves $\tau \cdot v_e$ per round trip has little importance for a circular path—but for Sommerfeld's elliptic orbits it may influence the precession movement of the ellipses' main axes.